# Revisión sistemática de la literatura del desarrollo de América Latina en el periodo 2010-2021

Systematic review of development literature from Latin America between 2010-2021

Pedro Alfonso de la Puente Sierra[1]
Juan José Berdugo Cepeda[2]
María José Pérez Pacheco[3]

Resumen

El propósito de esta revisión sistemática es identificar y describir el estado de la literatura del desarrollo en América Latina en español e inglés desde 2010. Para esto, realizamos una revisión topográfica de 44 artículos disponibles en los índices bibliográficos más importantes de América Latina, publicados en revistas de diversas disciplinas. Nuestro análisis se enfocó en analizar la naturaleza y la composición de la literatura, y encontró una gran proporción de artículos provenientes de revistas de México y Colombia, así como especializadas en la disciplina económica. Los artículos revisados más relevantes muestran diversidad metodológica y temática, con especial atención al problema del crecimiento en el desarrollo latinoamericano. Una limitación importante de esta revisión es la exclusión de artículos publicados en portugués y de literatura no indexada (como tesis y disertaciones). Esto conduce a diversas recomendaciones para revisiones futuras sobre la literatura del desarrollo producida en América Latina.

[1] Doctorando en Ciencias Sociales, Universidad del Norte, Colombia. Correo: pdelapuente@uninorte.edu.co
[2] Estudiante de Ciencias Políticas, Universidad del Norte, Colombia.
[3] Estudiante de Ciencias Políticas, Universidad del Norte, Colombia.

Abstract

The purpose of this systematic review is to identify and describe the state of development literature published in Latin America, in Spanish and English, since 2010. For this, we carried out a topographic review of 44 articles available in the most important bibliographic indexes of Latin America, published in journals of diverse disciplines. Our analysis focused on analyzing the nature and composition of literature, finding a large proportion of articles coming from Mexico and Colombia, as well as specialized in the economic discipline. The most relevant articles reviewed show methodological and thematic diversity, with special attention to the problem of growth in Latin American development. An important limitation of this review is the exclusion of articles published in Portuguese, as well as non-indexed literature (such as theses and dissertations). This leads to various recommendations for future reviews of the development literature produced in Latin America.

1. Introducción

El volumen de la literatura del desarrollo en América Latina tiene una historia tan larga como la historia de esta disciplina. Desde finales de la década de 1940, surgen los análisis que cuestionan las causas, las razones y los factores del desarrollo, así, en 1949, Raul Prebisch publicaba su ensayo seminal *El desarrollo económico de la América Latina y algunos de sus principales problemas*, que, además de ser uno de los primeros trabajos del desarrollo latinoamericano, también fue una obra pionera de la escuela del pensamiento estructuralista.

Desde entonces, se ha acumulado un corpus con gran variedad de perspectivas epistemológicas, metodológicas y temáticas, que alcanzaron su punto culminante en la década de 1970. Desde entonces, diversos autores han identificado un descenso continuo de la literatura del desarrollo (Hidalgo-Capitán, 2011; Hirschman, 1980). Sin embargo, en años recientes, marcados por el auge de los estudios micro del desarrollo, se habla del renacimiento de la disciplina, a través de la cooptación (Herrera, 2006) o del surgimiento de alternativas intelectuales (Escobar, 2007).

Esta afirmación nos motivó a revisar el estado de la literatura del desarrollo en América Latina para identificar la naturaleza de las publicaciones y la composición metodológica de estas.

Además, al buscar en índices y bases de datos bibliográficas como Scielo, Redalyc, Dialnet y Google Scholar artículos que contengan, ya sea en el título, ya sea en las palabras clave, los términos "revisión sistemática", "desarrollo económico" y "América Latina", no se arroja ningún resultado que analice la producción intelectual proveniente de la región sobre el desarrollo.[4] En este sentido, se hace patente la necesidad de realizar una revisión sistemática sobre el estado de la literatura reciente del desarrollo en América Latina.

Así, el propósito de la revisión es identificar y sintetizar el estado de la literatura, esto es, de la producción de conocimiento, de la investigación realizada desde América Latina sobre el desarrollo de la región, así como establecer su naturaleza bibliométrica y composición metodológica. Para dar cumplimiento, se plantean las siguientes preguntas de investigación, a saber:

- ¿Cuál es la naturaleza de la literatura sobre el desarrollo económico en América Latina desde el punto de vista del volumen, el impacto de citaciones y la distribución entre países, revistas y autores?
- ¿Cuál es la composición de la literatura relevante sobre el desarrollo en América Latina desde el punto de vista de los tipos de artículos, temas y métodos de investigación?

---

[4] Sí existen, no obstante, múltiples trabajos de revisión sistemática del desarrollo latinoamericano desde una perspectiva más limitada, esto es, revisando la literatura de los objetivos del milenio o el desarrollo sostenible.

Además, se definió un criterio de búsqueda y selección de artículos acorde con los requisitos de una revisión sistemática y un criterio de acotamiento para identificar la literatura relevante.

La estructura del documento se compone de la introducción, seguida de la sección metodológica en la que se describe el alcance de la revisión, los criterios de búsqueda y selección, así como el mecanismo de extracción y análisis de datos. La siguiente sección corresponde al análisis de resultados tanto de la naturaleza como de la composición de la literatura revisada y, por último, se discuten los resultados, las limitaciones e implicaciones de la revisión.

## 2. Metodología

### 2.1. Alcance de la revisión

Las revisiones de literatura son la base de la escritura científico-académica y por medio de estas el investigador se familiariza no solo con los artículos, sino que también identifica los autores y las publicaciones más relevantes (Trindade et al., 2017). Para dar cumplimiento a los objetivos planteados, adoptamos el método de revisión sistemática de la literatura del desarrollo en América Latina (Hallinger, 2013; Petticrew y Roberts, 2006).

Para la correcta realización de una revisión sistemática, su alcance debe identificar claramente el enfoque temático, geográfico, el periodo temporal y el tipo de fuentes (Castillo y Hallinger, 2018).

Nuestro enfoque temático consistió en artículos cuyo objetivo sea el estudio del desarrollo en América Latina, establecidos en sus palabras clave o títulos, publicados en español o inglés. La decisión de enfocarnos en publicaciones en español e inglés, se basó en diversos factores. Primero, el interés por identificar la literatura sobre el desarrollo en el idioma más hablado de la región, pues el 60 % de los 650 millones de latinoamericanos usamos el español como primera lengua, y al mismo tiempo incluir la literatura publicada en América Latina en el idioma internacional más aceptado para publicaciones científicas, pues al menos el 75 % de la literatura científica se publica en este idioma (Slate, 2015). En este sentido, el objetivo es incluir la mayor cantidad posible de artículos publicados en América Latina sobre el desarrollo.

La principal debilidad de esta decisión fue excluir los estudios publicados en portugués, principalmente debido a la falta de competencia en el lenguaje de los autores. En este sentido, reconocemos, desde el punto de partida, que esta revisión no cumple con ser una "revisión comprensiva de la literatura publicada sobre el desarrollo en América Latina".

El alcance geográfico de las publicaciones se definió como "América Latina", la cual incluye todos los países donde el español y el portugués son oficiales. Se determina que el inicio de la revisión es 2010, para incluir solo la literatura más reciente y conocer el estado de esta. El punto final de la revisión se estableció en marzo de 2021.

El criterio de búsqueda de una revisión puede clasificarse como selectivo, limitado o exhaustivo;[5] en este caso, realizamos una búsqueda limitada de fuentes. En este tipo de búsqueda, el autor delimita la revisión utilizando un criterio definido explícitamente, y cumple los requisitos de una revisión sistemática cuando tal criterio es claro y defendible (Hallinger, 2013). Nuestro criterio de búsqueda se definió como conjunto de publicaciones que, cumpliendo los criterios de enfoque temático, alcance geográfico y periodo temporal, se encuentren en los índices bibliográficos Scielo y Redalyc. La selección de estos dos índices se hace a partir de su importancia en el sistema de publicaciones académicas de América Latina.

---

[5] Hallinger (2013) establece que la búsqueda selectiva, en que no se establece el criterio claramente y resulta del juicio de los autores, no cumple los requisitos de una revisión sistemática de la literatura. Las búsquedas exhaustivas, por su parte, incluyen una amplia variedad de opciones de búsqueda, por ejemplo, múltiples bases de datos, palabras clave, métodos y revistas científicas; y en estas los autores deben realizar un trabajo más complejo, por lo que el uso de *software* especializados es recomendable.

Scielo (Scientific Electronic Library Online) es una base de datos de referencias a artículos publicados en más de 1000 revistas de acceso abierto de 12 países,[6] con cobertura desde 2002, de todas las áreas de conocimiento y con registros publicados en español, portugués e inglés.

Por su parte, Redalyc (Red de Revistas Científicas de América Latina y el Caribe, España y Portugal) fue creada en 2003, y es un sistema de indexación que integra revistas de alta calidad científica y editorial de la región, así como exclusivamente a las que comparten el modelo de publicación sin fines de lucro, para conservar la naturaleza académica y abierta de la comunicación científica; en 2019, incluía 1310 revistas con 50 000 fascículos y 650 000 artículos.

Así, el criterio básico de búsqueda y selección se puede resumir de la siguiente forma:

> Recuperar de los índices Scielo y Redalyc aquellos artículos, escritos en español o inglés, publicados entre enero de 2010 y marzo de 2021, en revistas latinoamericanas que contengan los términos "desarrollo" o "América Latina" en sus palabras clave, complementados por aquellos que, sin tenerlas, mencionen ambos términos en su título.

Del procedimiento de búsqueda se recuperaron títulos, autores, revistas, resúmenes y códigos JEL (Journal of Economic Literature) (donde estaban disponibles), para corroborar que tratan sobre el "desarrollo de América Latina".

2.2. Origen y análisis de datos e información

Durante el proceso de búsqueda, cuando se identificó una fuente relevante, esta se descargó y se revisaron datos tales como nombre y filiación de los autores, nombre de la revista, año de publicación, dominio geográfico, tipo de artículo, método de investigación y tema. Los datos relacionados se agregaron en una base de datos en MS Excel.

---

[6] Estos países son Argentina, Brasil, Chile, Colombia, Costa Rica, Cuba, España, México, Perú, Portugal, Sudáfrica y Venezuela.

Para el análisis de datos, realizamos unas estadísticas descriptivas para sintetizar los resultados e identificar los patrones de la literatura del desarrollo en América Latina.

3. Resultados

Se discuten los resultados de la revisión realizada, y dado que el objetivo es describir la producción de conocimiento sobre el desarrollo económico en América Latina, se buscan resolver las siguientes preguntas:

- ¿Cuál es la naturaleza de la literatura sobre el desarrollo económico en América Latina desde el punto de vista del volumen, el impacto de citaciones y la distribución entre países, revistas y autores?
- ¿Cuál es la composición de la literatura relevante sobre el desarrollo en América Latina desde el punto de vista de los tipos de artículos, temas y métodos de investigación?

Para dar respuesta a estas preguntas, se divide el análisis en dos partes, a saber:

Una discusión de los resultados para los 44 artículos que se identificó cumplen el criterio de búsqueda y selección. Así, para dar respuesta a la primera pregunta, la discusión incluye la información bibliométrica de la investigación, tal como país de origen de los autores y revistas, frecuencia de las revistas, palabras clave, codificación del JEL y citaciones.

Consecuentemente, para dar respuesta a la segunda pregunta, en particular al criterio relevante, se toma el número de citaciones anuales por artículo como factor definitivo de la relevancia. Así, aquellos que presenten un valor superior al promedio de citaciones anuales serán analizados, lo cual resulta en 13 artículos de los 44 identificados originalmente.

3.1. Resultados: naturaleza de la investigación

De la búsqueda en las bases de datos seleccionadas, Scielo y Redalyc, se toman 44 artículos, los cuales cuentan con 70 autores y fueron publicados en 31 revistas.

De los 70 autores registrados en la base de datos, 21 están afiliados a instituciones mexicanas y 9 a instituciones ecuatorianas. La mayoría de los autores están afiliados a instituciones latinoamericanas, aunque hay una importante presencia de autores en España y otros países por fuera de América (figura 1).

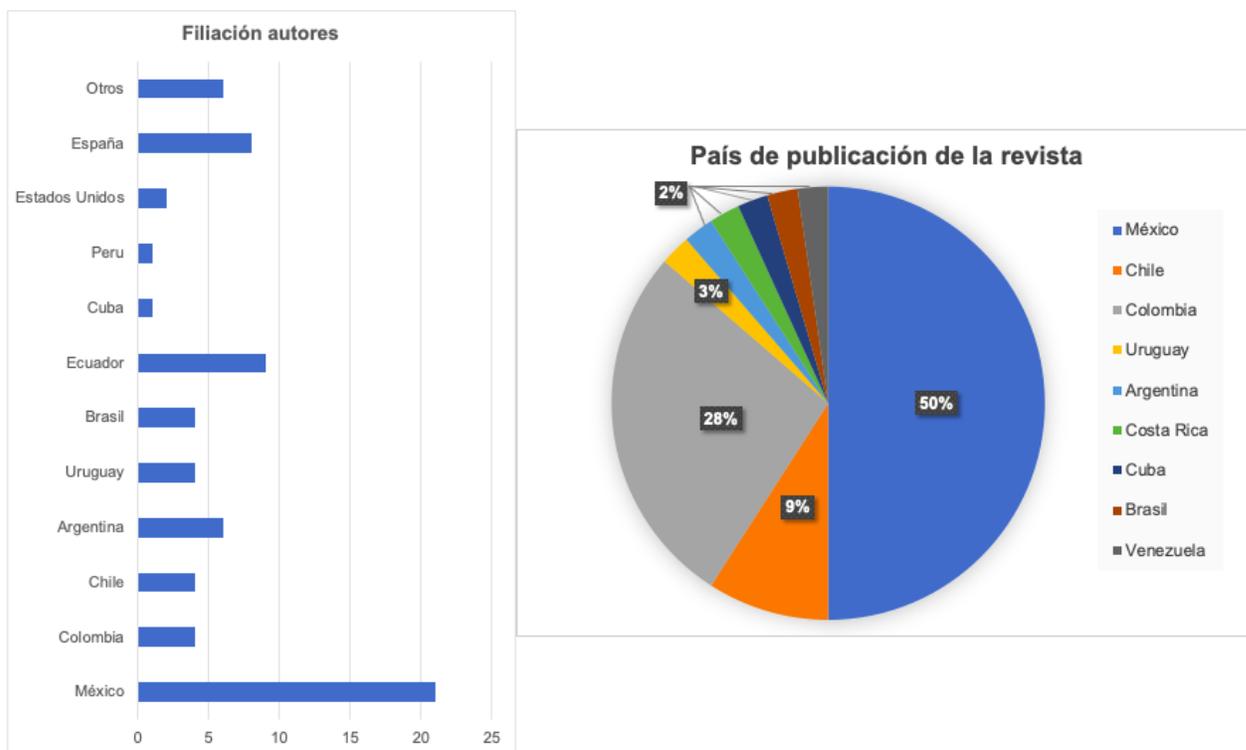

Figura 1. País de origen de los autores y revistas (n = 44)

Siguiendo el criterio de selección de artículos para la revisión, todos provienen de revistas latinoamericanas, pues Scielo y Redalyc solo indexan revistas de la región; en atención a que se seleccionaron artículos en español e inglés, hay un sesgo a subrepresentar la producción científica brasileña. En cualquier caso, de las 31 revistas, el 87 % de estas son publicadas por instituciones ubicadas en México (50 %), Colombia (28 %) y Chile (9 %).

Así, 22 de los 44 artículos fueron publicados en revistas mexicanas (Anser et al., 2020; Beteta y Moreno-Brid, 2012; Bresser-Pereira, 2017; Brugger y Ortiz, 2012; Cordera Campos, 2014; Domínguez y Caria, 2018; Favila, 2019; Hernández-Veleros, 2016; Hernández, 2017; Jaimes y Matamoros, 2017; López, 2020; Mballa, 2017; Morales, 2010; Nudelsman, 2013; Ornelas, 2012; Osorio, 2019; Osorio, 2012; Ros, 2011; Salama, 2020; Suanes y Roca-

Sagalés, 2015; Vernengo, 2020; Yetano y Castillejos, 2019), 12 en revistas colombianas (Cardona, 2020; Concha y Taborda, 2014; García, 2018; González, 2012; Iglesias y Carmona, 2016; Lara et al., 2018; Pinilla et al., 2013; Quinde et al., 2020; Resico, 2019; Ríos-Ávila, 2017; Rojas y Ramírez, 2018; Alarco, 2017), 4 en revistas chilenas (Bossio, 2020; Lehnert y Carrasco, 2020; Morales, 2012; Valenzuela, 2019), mientras que revistas de Argentina (Escobar, 2017), Brasil (Brida et al., 2021), Costa Rica (Levalle, 2018), Cuba (Kogan y Bondorevsky, 2016), Uruguay (Soto y Cerrano, 2020) y Venezuela (Quinde-Rosales et al., 2019) están representadas con un artículo.

Al revisar la frecuencia de publicación de las revistas (figura 2), se observa que *Problemas del Desarrollo* es la revista más común, con 7 artículos revisados (Brugger y Ortiz, 2012; Jaimes y Matamoros, 2017; López, 2020; Morales, 2010; Nudelsman, 2013; Ornelas, 2012; Vernengo, 2020) (c. 16 %), lo cual es consistente, pues el alcance de la publicación es divulgar las "interpretaciones teóricas que con rigor científico pretendan analizar las diferentes dificultades planteadas por el desarrollo económico"; también destaca *Economía UNAM*, la cual tiene un alcance más amplio (temas económicos en general) y está representada con 4 artículos (Beteta y Moreno-Brid, 2012; Bresser-Pereira, 2017; Cordera, 2014; Ros, 2011) (c. 10 %). Estas dos revistas, que concentran 11 de los 44 artículos revisados, son editadas por el Instituto de Investigaciones Económicas (IIEC) de la Universidad Nacional Autónoma de México (UNAM).

La tercera revista más frecuente, *El Trimestre Económico*, es editada por el Fondo de Cultura Económica (FCE), y tiene en común con las revistas de la UNAM la alta frecuencia temporal de sus números (trimestral y cuatrimestral), con más de 10 artículos por número en promedio.

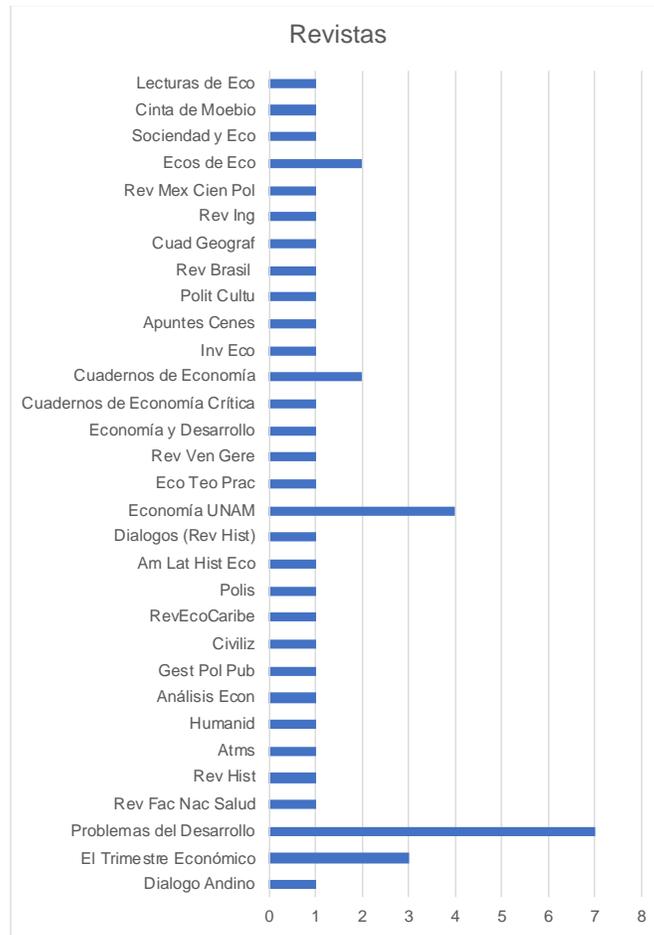

Figura 2. Título y frecuencia de las revistas (n = 44)

Ahora, dado el objetivo y alcance de la revisión, y en atención que el criterio de selección principal fue artículos publicados, en español o inglés, en revistas latinoamericanas que contengan las palabras "desarrollo" o "América Latina" en sus palabras clave, complementados por aquellos que, sin tenerlas, mencionen ambos términos en su título", no sorprende que estas sean las palabras clave más comunes (figura 3).

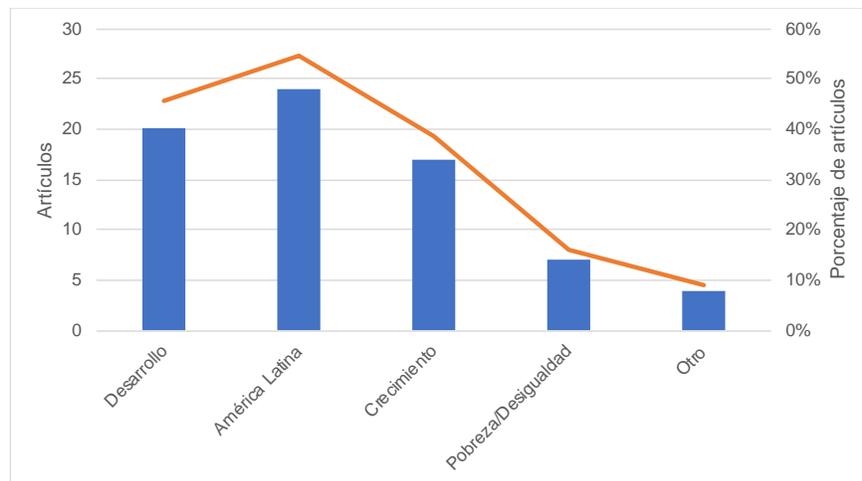

Figura 3. Principales palabras clave por artículo (n = 44)

Lo que sí resalta es la alta frecuencia del término "crecimiento", especialmente al compararla con los términos "pobreza/desigualdad", lo que indica, en primera instancia, que la literatura del desarrollo está más orientada a estudiar el problema del crecimiento en el desarrollo en lugar del problema de la desigualdad en el desarrollo, lo cual puede indicar asuntos epistémico-metodológicos. En cualquier caso, todas las palabras clave relacionan el desarrollo con una variable de interés de carácter macro, y no se observan palabras clave que asocien el desarrollo a temas actualmente relevantes, como los estudios de género, la experimentación causal, la economía del comportamiento o la ciencia de datos.

Además, uno de los criterios de búsqueda inicialmente considerados fue identificar los artículos con código JEL que registraran el código O54, correspondiente a *Desarrollo económico - Estudios de países: América Latina; Caribe*. Sin embargo, la búsqueda se amplió debido a:

- La relativa escasez de artículos en español (e inglés) que utilizan este código, lo cual de por sí es indicativo del estado de la literatura.

- La naturaleza multi- e interdisciplinar de la literatura del desarrollo, lo que implica artículos publicados en revistas que, al no ser de economía, no registran código JEL.

Esto último se nota en que, de los 44 artículos revisados, 22 no tienen código JEL asignado (figura 4).

Si bien nueve artículos sí registran el código JEL O54, es importante señalar que concurrentemente 16 registran un código distinto, pero contenido dentro de la categoría *O - Desarrollo económico, innovación, cambio tecnológico y crecimiento*.[7]

La justificación de este fenómeno puede estar asociada a que el estudio del (problema del) desarrollo en América Latina permite resultados/observaciones acerca del desarrollo sin necesariamente ser un estudio regional o de país.

Por fuera del código O, hay una amplia variedad de códigos asociados, entre ellos los F (Economía internacional) y N (Historia económica).

---

[7] En particular, O1 - Desarrollo económico; O2 - Políticas y planeación del desarrollo; O4 - Crecimiento económico y productividad agregada; O5 - Estudios de países. Además, es importante añadir que ninguno de los artículos revisados con códigos JEL incluye alguno de la categoría O3 - Innovación, investigación y desarrollo, cambio tecnológico, derechos de propiedad intelectual.

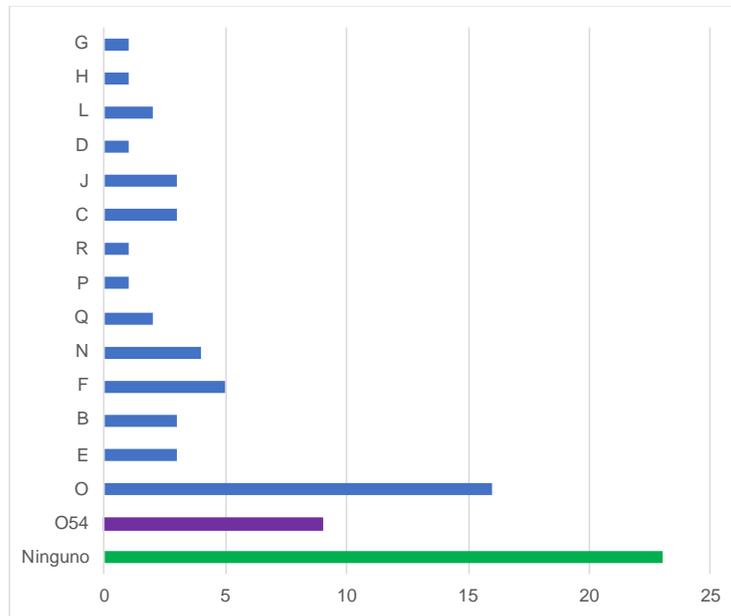

Figura 4. Frecuencia de códigos JEL en artículos (n = 44)

Ahora, para revisar el número de citaciones de los artículos, utilizamos la herramienta Publish or Perish de Harzing, en la cual se realizó una búsqueda de citaciones a través de Google Scholar. Se obtuvieron los siguientes resultados (figuras 5-6; anexo 1):

- De los 44 artículos, 33 tienen citaciones mayores de uno, mientras 11 tienen cero citaciones
- De los 11 artículos con cero citaciones:
    - Cinco fueron publicados en 2019
    - Cinco en 2020
    - Uno en 2021
- Al revisar las citaciones para los 33 artículos con más de una citación, el promedio de citaciones de estos es de 3,1 citas anuales, mientras el valor mediano es de 2,3 citas anuales.

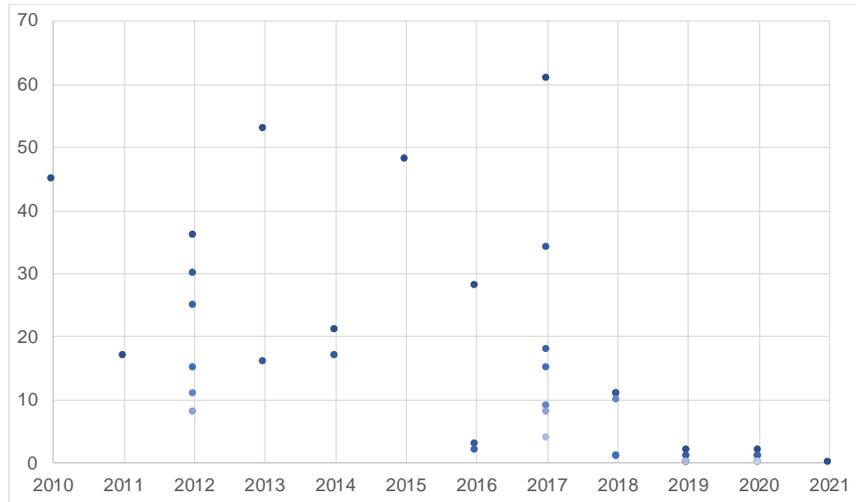

Figura 5. Total citas por artículo y por año de publicación (n = 44)

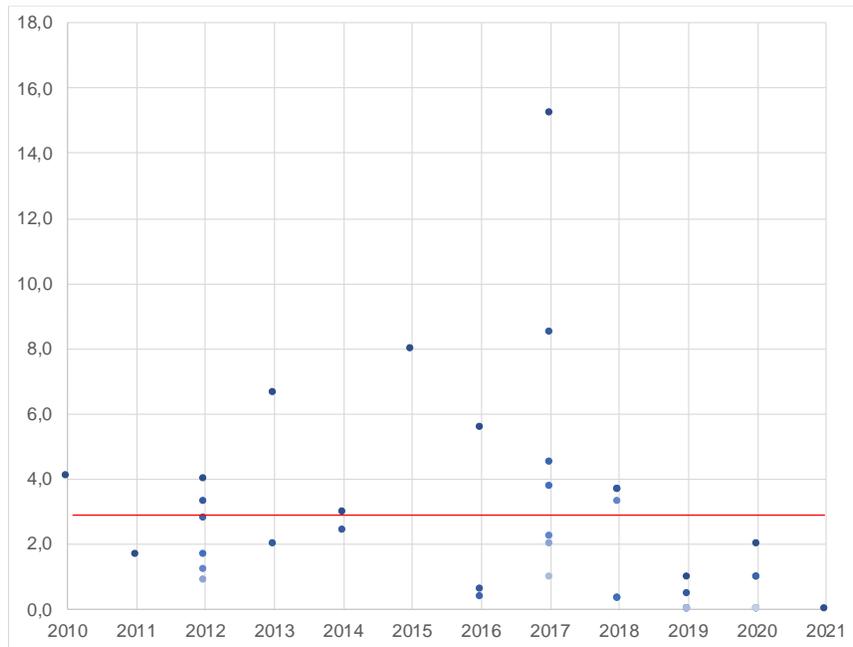

Figura 6. Citas anuales por artículo (n = 44)

Consistente con lo señalado, la mayoría de los artículos con citaciones anuales inferiores al promedio (3,1 citas anuales) se concentran en el periodo de publicación más reciente, pues todos los artículos publicados desde 2019 tienen un promedio de citaciones anual de 0,34 citas.

Sin embargo, también destaca que, de los 13 artículos revisados que fueron publicados en 2015 o antes, ocho presentan un valor promedio de citaciones inferior al promedio anual,[8] lo que nos indica que no toda la literatura publicada en español o inglés en revistas latinoamericanas sobre el desarrollo de la región es relevante.

3.2.    Resultados: composición de la literatura

En atención al criterio de relevancia anotado en la pregunta de investigación, se decidió acotar la selección de artículos, para incluir solo aquellos que cumplan el criterio de relevancia definido.

Así, al ser la frecuencia de citaciones la base del análisis de la composición de la literatura, se decidió acotar el número de artículos revisados utilizando como marco de referencia para la *relevancia* los artículos que posean un valor de citaciones anuales superior al promedio de los 44 artículos iniciales, correspondiente a 3,1, de cual resulta que 13 artículos cumplieron con este criterio. Vale aclarar que la justificación para la selección de este marco es arbitraria y se establece para acotar el análisis. Así, los artículos son los siguientes:

Tabla 1. Descripción de los artículos (n = 13)

| | Título | Autores | Citas | Citas por año | Año | Revista |
|---|---|---|---|---|---|---|
| 1 | "Desarrollo local y microfinanzas como estrategias de atención a las necesidades sociales: Un acercamiento teórico conceptual" | Louis Valentin Mballa | 61 | 15,25 | 2017 | *Revista Mexicana de Ciencia Política* |
| 2 | "La nueva teoría desarrollista: Una síntesis" | Luiz Carlos Bresser-Pereira | 34 | 8,50 | 2017 | *Economía UNAM* |
| 3 | Inversión extranjera directa, crecimiento económico y desigualdad en América Latina | Macarena Suanes y Oriol Roca-Sagalés | 48 | 8,00 | 2015 | *El Trimestre Económico* |

---

[8] Estos ocho artículos promedian las 1,96 citas anuales.

| # | Título | Autores | | | Año | Revista |
|---|--------|---------|---|---|------|---------|
| 4 | "Gasto público y crecimiento económico: Un estudio empírico para América Latina" | Diego Pinilla Rodríguez, Juan Jiménez Aguilera y Roberto Montero Granados | 53 | 6,63 | 2013 | *Cuadernos de Economía* |
| 5 | "La infraestructura en el desarrollo de América Latina" | Jorge Kogan y Diego Bondorevsky | 28 | 5,60 | 2016 | *Economía y Desarrollo* |
| 6 | "Ciclos distributivos y crecimiento económico en América Latina, 1950-2014" | Germán Alarco Tosoni | 18 | 4,50 | 2017 | *Cuadernos de Economía* |
| 7 | "Inversión extranjera directa y desarrollo en América Latina" | Josefina Morales | 45 | 4,09 | 2010 | *Problemas del Desarrollo* |
| 8 | "Volver al desarrollo" | Jaime Ornelas Delgado | 36 | 4,00 | 2012 | *Problemas del Desarrollo* |
| 9 | "Unions and economic performance in developing countries: Case studies from Latin America" | Fernando Ríos-Ávila | 15 | 3,75 | 2017 | *Ecos de Economía* |
| 10 | Materias primas críticas y complejidad económica en América Latina | Juan Sebastián Lara-Rodríguez, André Tosi Furtado y Aleix Altimiras-Martin | 11 | 3,67 | 2018 | *Apuntes del Cenes* |
| 11 | Inversión en infraestructura vial y su impacto en el crecimiento económico: Aproximación de análisis al caso infraestructura en Colombia (1993-2014) | Miguel David Rojas-López y Andrés Felipe Ramírez-Murie | 11 | 3,67 | 2018 | *Revista de Ingeniería Universidad de Medellín* |
| 12 | "Mercados accionarios y su relación con la economía real en América Latina" | Samuel Brugger y Edgar Ortiz | 30 | 3,33 | 2012 | *Problemas del Desarrollo* |
| 13 | Raíces latinoamericanas del otro desarrollo: Estilos de desarrollo y desarrollo a escala humana" | Rafael Domínguez y Sara Caria | 10 | 3,33 | 2018 | *América Latina en la Historia Económica* |

Sobre estos 13 artículos podemos decir que:

- o Diez están publicados en revistas especializadas en economía
- o Siete están publicados en revistas mexicanas, cinco en colombianas y una revista cubana

- Ocho contienen código JEL y cuentan con, al menos, un código correspondiente a la categoría *O - Desarrollo económico, innovación, cambio tecnológico y crecimiento.*

Ahora, de esta selección acotada de artículos, compuesta por los más citados, tenemos que se pueden clasificar en tres tipos de métodos (tabla 2; figura 7), a saber:

- Artículos de revisión, los cuales presentan el panorama de la literatura que relaciona el desarrollo con una variable de interés, tal como el desarrollo local o la inversión extranjera. Si bien en la selección solo hay tres de estos (Domínguez y Caria, 2018; Mballa, 2017; Rojas y Ramírez, 2018), uno corresponde al articulo más citado (Mballa, 2017), casi duplicando en citaciones al siguiente (15,25 vs. 8m5).
- Artículos de reflexión, los cuales, según el reconocimiento de los autores en el campo, tienen más o menos citas, destacando por su carácter propositivo/crítico de la disciplina del desarrollo; de los 13 artículos, cuatro corresponden a este tipo (Bresser-Pereira, 2017; Kogan y Bondorevsky, 2016; Morales, 2010; Ornelas, 2012).
- Artículos científicos con análisis estadístico: en la revisión, se encontraron diversos documentos con gran número de citas, cuyo elemento central es la prueba de hipótesis usando métodos econométricos; en particular, cinco de los 13 artículos se pueden categorizar (Brugger y Ortiz, 2012; Pinilla et al., 2013; Ríos-Ávila, 2017; Suanes y Roca-Sagalés, 2015; Alarco, 2017), mientras otros dos realizan análisis estadísticos menos complejos (Lara et al., 2018; Rojas y Ramírez, 2018). Un detalle interesante de estos artículos es que, si bien contienen menos elementos críticos que los artículos de revisión y, naturalmente, que los de reflexión, sí parecen reconocer las limitaciones del análisis estadístico e intentan ofrecer una visión/interpretación amplia de la relación entre el desarrollo y su variable de interés.

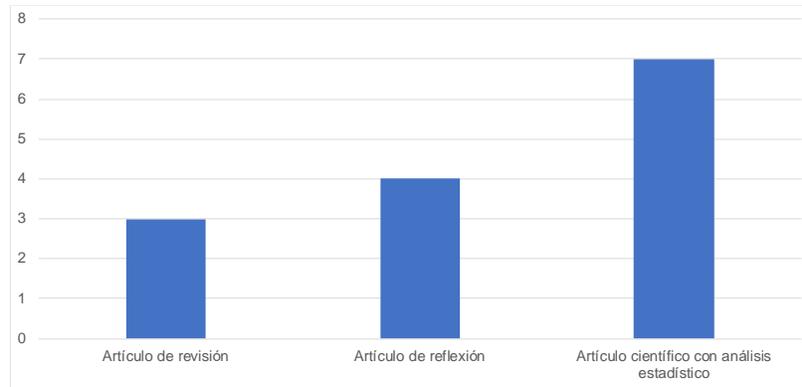

Figura 7. Tipología de los artículos revisados (n = 13)

De la lista de 13 artículos, dos contienen métodos difíciles de categorizar, a saber: uno corresponde al método de evaluación comparativa, que incluimos en la categoría *artículo científico con análisis estadístico*, pues, aunque no corresponde a análisis econométrico o descriptivo, sí realiza cálculos estadísticos (Lara et al., 2018); y otro realiza un método mixto de revisión y análisis estadístico, por lo que lo incluimos en ambas categorías (Rojas y Ramírez, 2018).

Tabla 2. Métodos de los artículos (n = 13)

| | Artículo | Método | Tipo de artículo y método descrito |
|---|---|---|---|
| 1 | Mballa, L. V. (2017). Desarrollo local y microfinanzas como estrategias de atención a las necesidades sociales: Un acercamiento teórico conceptual. *Revista Mexicana de Ciencias Políticas y Sociales*, 62(229), 101-127. | Cualitativo | Artículo de revisión: revisión analítica de la literatura |
| 2 | Bresser-Pereira, L. C. (2017). La nueva teoría desarrollista: Una síntesis. *Economía UNAM*, 14(40), 48-66. | Cualitativo | Artículo de reflexión: síntesis histórico-analítica del desarrollo |
| 3 | Suanes, M. y Roca-Sagalés, O. (2015). Inversión extranjera directa, crecimiento económico y desigualdad en América Latina. *El Trimestre Económico*, 82(327), 675-706. | Cuantitativo | Artículo científico con análisis econométrico: panel de datos para 18 países entre 1980 y 2009, que usa de modelos de EF, MC2E y MGM. |
| 4 | Pinilla Rodríguez, D. E., Jiménez Aguilera, J. de D. y Montero Granados, R. (2013). Gasto público y crecimiento económico: Un estudio empírico para América Latina. *Cuadernos de Economía*, *32*(59), 179-208. | Cuantitativo | Artículo científico con análisis econométrico: panel de datos para 17 países entre 1989 y 2009, que usa modelos de MCO combinados no lineales y MCG de efectos fijos y variables. |
| 5 | Kogan, J. y Bondorevsky, D. (2016). La infraestructura en el desarrollo de América Latina. *Economía y Desarrollo*, *156*(1), 168-186. | Cualitativo | Artículo de reflexión: estado de la cuestión. |
| 6 | Alarco Tosoni, G. (2017). Ciclos distributivos y crecimiento económico en América Latina, 1950-2014. *Cuadernos de Economía*, *36*(spe72), 1-42. | Cuantitativo | Artículo científico con análisis econométrico: panel de datos, que usa de modelos asociados (EF, MV y Arellano-Bond). |
| 7 | Morales, J. (2010). Inversión extranjera directa y desarrollo en América Latina. *Problemas del Desarrollo*, *41*(163), 141-156. | Cualitativo | Artículo de reflexión: estado de la cuestión. |
| 8 | Ornelas Delgado, J. (2012). Volver al desarrollo. *Problemas del Desarrollo*, *43*(168), 7-35. | Cualitativo | Artículo de reflexión: síntesis histórico-analítica del desarrollo. |
| 9 | Ríos-Ávila, F. (2017). Unions and economic performance in developing countries: Case studies from Latin America. *Ecos de Economía*, *21*(44), 4-36. | Cuantitativo | Artículo científico con análisis econométrico: panel de datos para seis países, que usa de modelo de Brown y Medoff (variación Cobb-Douglas). |
| 10 | Lara-Rodríguez, J. S., Furtado, A. T. y Altimiras-Martin, A. (2018). Materias primas críticas y complejidad económica en América Latina. *Apuntes del Cenes*, *37*(65), 15-51. | Cuantitativo | Método de evaluación comparativo de complejidad económica. |
| 11 | Rojas-López, M. D. y Ramírez-Muriel, A. F. (2018). Inversión en infraestructura vial y su impacto en el crecimiento económico: Aproximación de análisis al caso infraestructura en Colombia (1993-2014). *Revista Ingenierías Universidad de Medellín*, *17*(32), 109-128. | Mixto | Artículo de revisión y científico: incluye revisión de literatura y análisis estadístico de interrelaciones (regresión lineal simple). |
| 12 | Brugger, S. y Ortiz, E. (2012). Mercados accionarios y su relación con la economía real en América Latina. *Problemas del Desarrollo*, *43*(168), 63-93. | Cuantitativo | Artículo científico con análisis econométrico: análisis de series de tiempo, que usa siete modelos econométricos. |
| 13 | Domínguez, R. y Caria, S. (2018). Raíces latinoamericanas del otro desarrollo: Estilos de desarrollo y desarrollo a escala humana. *América Latina en la Historia Económica*, *25*(2), 175-209. | Cualitativo | Artículo de revisión: reconstrucción histórica. |

Metodológicamente, los artículos científicos con análisis estadístico muestran una concentración en análisis de tipo panel de datos con modelos asociados, tales como de efectos fijos (EF), máxima verosimilitud (MV) y método generalizado de momentos (MGM), entre los más usados Los artículos de revisión y reflexión no ofrecen un apartado para describir la metodología, pero por su naturaleza no lo requieren. Los objetivos de los artículos (anexo 2) son variados y dependen, principalmente, de la tipología del artículo; así, los artículos científicos con análisis estadístico buscan probar hipótesis que relacionan sus variables de interés con el crecimiento o desarrollo económico de América Latina, mientras que los artículos de revisión y reflexión buscan ofrecer un panorama crítico de la situación económica y del desarrollo en la región.

En cualquier caso, esta tipología nos permite ubicar la literatura relevante por métodos de investigación, teniendo que los seis artículos de revisión y reflexión se pueden ubicar en el método *cualitativo*, pues la observación y el análisis documental (insumos centrales de este tipo de artículos) se ubican en el método, mientras que seus artículos científicos con análisis estadístico lo hacen en el método *cuantitativo*, y uno en el método mixto.

4. Discusión y conclusiones

En esta revisión sistemática, se documentó el estado de la investigación publicada desde América Latina sobre el desarrollo de la región en el periodo 2010-2021. Sintetizamos la naturaleza de los 44 artículos que cumplieron el criterio de búsqueda y selección, y se analizó la composición de los 13 artículos más relevantes.

En esta sección, se resumen algunas de las interpretaciones de los resultados principales, se resaltan las limitaciones del artículo y las implicaciones para investigaciones futuras.

### 4.1. Resumen e interpretación de resultados

El primer resultado de interés corresponde al volumen de la producción científica publicada desde América Latina sobre el desarrollo de la región. Así, identificamos 44 artículos publicados en 21 revistas de diversas disciplinas, entre ellas predominan las revistas de economía. Aunque la importancia de la disciplina económica es destacada, la multi- e interdisciplinariedad de los estudios del desarrollo resalta en que el 50 % de estos artículos fueron publicados en revistas que no incluyen códigos JEL en sus requisitos, el cual es el identificador más ampliamente usado de la literatura económica.[9]

También destaca que, en la revisión de los 44 artículos, el término que más frecuentemente se asocia en los artículos es "crecimiento", especialmente al compararlo con los términos "pobreza" o "desigualdad", lo que indica, en primera instancia, que la literatura del desarrollo está más orientada a estudiar el problema del crecimiento en el desarrollo, en lugar del problema de la desigualdad en el desarrollo. Esto indica asuntos epistémico-metodológicos, ya que todos los artículos que contienen "crecimiento" en sus palabras clave están publicados en revistas especializadas en economía, mientras que los que contienen "pobreza" o "desigualdad" fueron publicados en revistas con alcances muy amplios y de diversas disciplinas (*i. e.*, economía, ciencia política, salud pública, ciencias sociales).

De mismo modo, resalta que todas las palabras clave de los artículos relacionan el desarrollo con una variable de interés de carácter macro, y no se observan palabras clave que asocien el desarrollo a temas actualmente relevantes como estudios de género, experimentación causal, economía del comportamiento o ciencia de datos.

---

[9] Esto es, se espera que todas las revistas especializadas en economía asignen un código JEL a sus artículos, lo cual no es necesario para revistas de otras disciplinas.

Por otra parte, el IIEC de la UNAM es el principal medio de publicación para la academia latinoamericana del desarrollo, pues de sus revistas salieron el 25 % de los 44 artículos revisados. Es interesante que estos artículos presenten una alta variedad metodológica, incluso artículos de revisión, artículos de reflexión y artículos científicos con análisis estadístico. En cualquier caso, es notable el poder y la reputación que esta institución tiene sobre la academia latinoamericana sobre desarrollo.

Del análisis de citaciones, encontramos que la mayoría de los artículos de reciente publicación (desde 2018) tienen pocas o nulas citaciones, lo que indica: a) fallas o un importante rezago en la divulgación de las publicaciones sobre el desarrollo, b) debilidad general de la literatura del desarrollo en América Latina que como disciplina ha perdido relevancia o c) el cambio en la estructura de publicaciones sobre la disciplina del desarrollo hacia publicaciones en inglés no latinoamericanas.

Finalmente, al revisar los 13 artículos más relevantes, encontramos que existe una paridad en el método preferido de artículos. Así, seis se identifican como cualitativos, seis como cuantitativos y uno como mixto. De mismo modo, aunque los artículos científicos con análisis estadístico representan la mitad de los artículos revisados relevantes, los artículos de revisión y reflexión tienen mayor número de citas y citaciones por año, lo que resalta la importancia de la observación y revisión documental para el avance de la disciplina.

4.2. Limitaciones de la revisión

En la revisión, logramos identificar, al menos, tres limitaciones importantes. Una primera limitación es que, dado el propósito de la revisión, a saber: identificar y sintetizar las características de la producción de conocimiento, no se realiza un análisis de los resultados o las interpretaciones sustantivas de la investigación publicada en América Latina sobre el desarrollo, por lo cual no se pueden hacer interpretaciones sobre los patrones de producción intelectual ni sobre las fortaleza o debilidades de la literatura (Castillo y Hallinger, 2018).

La segunda limitación es que, siguiendo el criterio de búsqueda y selección, una parte importante de la literatura latinoamericana del desarrollo no está contenida en esta revisión. Así, por ejemplo, no se incluyen tesis de maestría o disertaciones doctorales, y no se observa ningún artículo proveniente de la *Revista Cepal*, la cual es ampliamente reconocida en el campo de los estudios del desarrollo económico, precisamente por no estar indexada tanto en Scielo como en Redalyc. Si bien es una limitación importante, la decisión de definir estos índices como base del criterio de búsqueda descansa en que incluye la mayoría de las publicaciones en español realizadas en revistas e instituciones latinoamericanas, siendo así las bases de datos más completas de la literatura en español disponibles.

Finalmente, la limitación final se relaciona con la decisión de excluir del criterio de búsqueda y selección los estudios publicados en portugués, principalmente debido a la falta de competencia en este idioma de los autores. En este sentido, reconocemos, desde el punto de partida que esta revisión no cumple con ser una revisión comprensiva de la literatura publicada sobre el desarrollo en América Latina.

Así, esta revisión solo busca servir como punto de partida para la ampliación del entendimiento de la literatura reciente publicada en América Latina sobre el desarrollo de la región.

### 4.3. Implicación de los resultados

La importancia de la revisión descansa en que es un esfuerzo importante por documentar el estado de la producción intelectual en la disciplina en América Latina, y a pesar de las limitaciones señaladas, nuestros resultados ofrecen algunas descripciones importantes. De este modo, se resalta la importancia de la disciplina en América Latina y legitima la necesidad de esta en la producción intelectual global del desarrollo.

Entre los diversos retos que se identificaron en esta revisión, destacan los siguientes: a) la necesidad de revisar la importancia del cambio estructural hacia publicaciones en inglés en revistas de alto impacto de instituciones ubicadas en Europa o América del Norte; b) identificar la existencia de un rezago intelectual, metodológico o divulgativo en la literatura del desarrollo latinoamericano, y c) reconocer si existe una debilidad general de la disciplina global y el estado de la literatura latinoamericana.

## Referencias

Alarco Tosoni, G. (2017). Ciclos distributivos y crecimiento económico en América Latina. *Cuadernos de Economía*, *36*(72), 1-42. https://doi.org/10.15446/cuad.econ.v36n72.65819

Anser, M. K., Hanif, I., Alharthi, M. y Chaudhry, I. S. (2020). Impact of fossil fuels, renewable energy consumption and industrial growth on carbon emissions in Latin American and Caribbean economies. *Atmosfera*, *33*(3), 201-213. https://doi.org/10.20937/ATM.52732

Beteta, H. y Moreno-Brid, J. C. (2012). El desarrollo en las ideas de la Cepal. *Economía UNAM*, *9*(27), 76-90. http://www.scielo.org.mx/scielo.php?pid=S1665-952X2012000300004&script=sci_abstract&tlng=pt

Bossio García, H. (2020). Clave hermenéutica del desarrollismo. *Revista de Historia*, *27*(1), 7-24. http://dx.doi.org/10.29393/rh27-2chg10002

Bresser-Pereira, L. C. (2017). La nueva teoría desarrollista: Una síntesis. *Economía UNAM*, *14*(40). http://www.scielo.org.mx/scielo.php?pid=S1665-952X2017000100048&script=sci_abstract&tlng=pt

Brida, J. G., Olivera, M. y Segarra, V. (2021). Economic growth and tourism performance in Latin America and the Caribbean: a comparative analysis by clustering techniques and causality tests. *Revista Brasileira de Pesquisa Em Turismo*, *15*(1), 1-15. https://doi.org/10.7784/rbtur.v15i1.2300

Brugger, S. y Ortiz, E. (2012). Mercados accionarios y su relación con la economía real en América Latina. *Problemas del Desarrollo*, *168*(43), 63-93. http://www.scielo.org.mx/scielo.php?pid=s0301-70362012000100004&script=sci_arttext

Cardona Arias, J. A. (2020). Evaluación del impacto económico de programas sociales contra la pobreza: Una revisión de estudios aleatorizados en la obra de Esther Duflo. *Revista de la Facultad Nacional de Salud Pública*, *38*(2). https://doi.org/10.17533/udea.rfnsp. e338856

Castillo, F. C. y Hallinger, P. (2018). Systematic review of research on educational leadership and management in Latin America, 1991-2017. *Educational Management Administration and Leadership*, *46*(2), 207-225. https://doi.org/10.1177/1741143217745882

Concha, Á. y Taborda, R. (2014). Insurance use and economic growth in Latin America: Some panel data evidence. *Lecturas de Economía*, *81*, 31-55. http://www.scielo.org.co/scielo.php?pid=S0120-25962014000200002&script=sci_arttext&tlng=en

Cordera Campos, R. (2014). El desarrollo ayer y hoy: Idea y utopía. *Economía UNAM*, *11*(33), 3-25. https://doi.org/10.1016/S1665-952X(14)72179-6

Domínguez, R. y Caria, S. (2018). Raíces latinoamericanas del otro desarrollo: Estilos de desarrollo y desarrollo a escala humana. *América Latina en la Historia Económica*, *25*(2), 175-209. https://doi.org/10.18232/alhe.898

Escobar, A. (2007). *La invención del tercer mundo*. El perro y la rana.

Escobar Váquiro, N. (2017). Avances fundamentales de la economía feminista en América Latina. *Cuadernos de Economía Crítica*, *4*(7), 17-41. https://ri.conicet.gov.ar/handle/11336/66980

Favila Tello, A. (2019). Eficiencia de la innovaciónen América Latina: Una aproximación a través del análisis envolvente de datos. *Análisis Económico*, *34*(87), 249-267. http://www.scielo.org.mx/scielo.php?pid=S2448-66552019000300249&script=sci_arttext

García, A. (2018). Rise and fall of regional planning in Argentina: Between the quest for embedded autonomy and the economic scenario of peripheral capitalism (1965-2015). *Cuadernos de Geografía: Revista Colombiana de Geografía*, *27*(1), 180-201. https://doi.org/10.15446/rcdg.v27n1.58053

González Molina, R. I. (2012). Desarrollo económico de América Latina y las integraciones regionales del siglo XXI. *Ecos de Economía*, *16*(35), 123-161. https://doi.org/10.17230/ecos.2012.35.6

Hallinger, P. (2013). A conceptual framework for systematic reviews of research in educational leadership and management. *Journal of Educational Administration*, *51*(2), 126-149. https://doi.org/10.1108/09578231311304670

Hernández-Veleros, Z. S. (2016). Modelos de crecimiento, estacionariedad y rompimientos: Comparación entre las tendencias de crecimiento de las economías de la OCDE y las de los países menos desarrollados. *El Trimestre Económico*, *83*(332), 635-678. https://doi.org/10.20430/ete.v83i332.235

Herrera, R. (2006). The neoliberal "rebirth" of development economics. *Monthly Review*, *58*(1), 38-50. https://doi.org/10.14452/MR-058-01-2006-05_4

Hidalgo-Capitán, A. L. (2011). Political economy of development: The retrospective construction of an academic branch. *Revista de Economía Mundial*, *28*, 279-320.

Hirschman, A. O. (1980). Auge y ocaso de la teoría económica del desarrollo. *El Trimestre Económico*, *47*(188), 1055-1077.

Hernández López, M. H. (2017). Variedades de capitalismo, implicaciones para el desarrollo de América Latina. *Economía: Teoría y Práctica*, *46*, 195-226. https://doi.org/10.24275/etypuam/ne/462017/hernandezlopez

Iglesias Lesaga, E. y Carmona Motolinia, J. R. (2016). Desigualdades, territorios y vulnerabilidades en el desarrollo de América Latina (1990-2015). *Revista Sociedad y Economía*, *31*, 123-148. https://doi.org/10.25100/sye.v0i31.3891

Jaimes, P. y Matamoros, G. (2017). ¿Qué sucedió con el casillero vacío del desarrollo de América Latina 30 años después? *Revista Problemas del Desarrollo*, *191*(48), 9-26. https://doi.org/10.1016/j.rpd.2017.11.002

Kogan, J. y Bondorevsky, D. (2016). La infraestructura en el desarrollo de América Latina. *Economía y Desarrollo*, *156*(1), 168-186. http://scielo.sld.cu/scielo.php?script=sci_arttext&pid=S0252-85842016000100012

Lara Rodríguez, J. S., Tosi Furtado, A. y Altimiras-Martin, A. (2018). Materias primas críticas y complejidad económica en América Latina. *Apuntes del Cenes*, *37*(65), 15-51. https://doi.org/10.19053/01203053.v37.n65.2018.5426

Lehnert, M. y Carrasco, N. (2020). del vivir bien y del desarrollo sustentable: Extractivismos y construcción de alternativas al desarrollo en Bolivia y Chile. *Diálogo Andino*, *63*, 189-204. http://dx.doi.org/10.4067/S0719-26812020000300189

Levalle, S. (2018). Recetas contra el conjuro: Los estilos de desarrollo en el Uruguay contemporáneo y los debates sobre el desarrollo latinoamericano (1973-2014). *Diálogos: Revista Electrónica de Historia*, *19*(1), 102-129. http://dx.doi.org/10.15517/dre.v19i1.30531

López, J. (2020). Raúl Prebisch y el pensamiento estructuralista latinoamericano. *Problemas del Desarrollo*, *51*(202), 3-24. https://doi.org/10.22201/iiec.20078951e.2020.202.69634

Mballa, L. V. (2017). Desarrollo local y microfinanzas como estrategias de atención a las necesidades sociales: Un acercamiento teórico conceptual. *Revista Mexicana de Ciencias Políticas y Sociales*, *62*(229), 101-127. http://www.scielo.org.mx/scielo.php?pid=S0185-19182017000100101&script=sci_abstract&tlng=pt

Morales, J. (2010). Inversión extranjera directa y desarrollo en América Latina. *Revista Problemas del Desarrollo*, *163*(41), 141-156. http://www.scielo.org.mx/scielo.php?pid=S0301-70362010000400008&script=sci_arttext

Morales, J. J. (2012). De los aspectos sociales del desarrollo económico a la teoría de la dependencia. *Cinta Moebio*, *45*, 235-252. https://doi.org/10.4067/S0717-554X2012000300004

Nudelsman, S. (2013). Implicaciones de la crisis financiera y económica global en América Latina. *Problemas del Desarrollo*, *175*(44), 125-146. http://dx.doi.org/10.4067/S0717-554X2012000300004

Ornelas Delgado, J. (2012). Volver al desarrollo. *Revista Problemas del Desarrollo*, *168*(43), 7-35. http://www.scielo.org.mx/scielo.php?pid=S0301-70362012000100002&script=sci_arttext

Osorio Caballero, M. I. (2019). Is Latin America's economic growth convergence procyclical? *Investigación Económica*, *78*(307), 33-53. https://doi.org/10.22201/fe.01851667p.2019.307.68446

Osorio, J. (2012). América Latina bajo el fuego de las grandes transformaciones económicas y políticas. *Política y Cultura*, *37*, 65-84. http://www.scielo.org.mx/scielo.php?pid=S0188-77422012000100004&script=sci_arttext

Petticrew, M. y Roberts, H. (2006). *Systematic reviews in the social sciences : A practical guide*. Blackwell Pub.

Pinilla Rodríguez, D. E., Jiménez Aguilera, J. y Montero Granados, R. (2013). Gasto público y crecimiento económico: Un estudio empírico para América Latina. *Cuadernos de Economía*, *35*(59), 181-210. http://www.scielo.org.co/scielo.php?script=sci_arttext&pid=S0121-47722013000100009

Quinde Rosales, V. X., Vaca Pinela, G., Quinde Rosales, F. y Lazo Vaca, L. (2020). Cointegration analysis between economic growth and environmental deterioration: An empirical analysis of sustainable development in Latin America and the Caribbean. *Revista de Economía del Caribe*, *24*, 8-25. https://doi.org/10.14482/ecoca.24.338.98

Quinde-Rosales, V., Bucaram-Leverone, R. y Bucaram-Leverone, M. (2019). Producto interno bruto en América Latina y el Caribe: Relaciones entre crecimiento económico y sustentabilidad ambiental. *Revista Venezolana de Gerencia*, *24*(87), 769-780. https://www.redalyc.org/articulo.oa?

Resico, M. F. (2019). Economía social de mercado versus capitalismo rentista: Reflexiones para América Latina. *Civilizar*, *19*(37), 103-116. https://doi.org/10.22518/usergioa/jour/ccsh/2019.2/a07

Ríos-Ávila, F. (2017). Unions and Economic Performance in developing countries: Case studies from Latin America. *Ecos de Economía*, *21*(44), 4-36. https://doi.org/10.17230/ecos.2017.44.1

Rojas López, M. D. y Ramírez Muriel, A. F. (2018). Inversión en infraestructura vial y su impacto en el crecimiento económico: Aproximación de análisis al caso infraestructura en Colombia (1993-2014). *Revista Ingenierías Universidad de Medellín*, *17*(32), 109-128. https://doi.org/10.22395/rium.v17n32a6

Ros, J. (2011). La productividad y el desarrollo en América Latina: Dos interpretaciones. *Economía UNAM*, *8*(23), 37-52. https://doi.org/10.22201/fe.24488143e.2011.23.149

Salama, P. (2020). Why do latin american countries suffer a long-term economic stagnation? A study from the cases of Argentina, Brazil and Mexico. *El Trimestre Económico*, *87*(348), 1083-1132. https://doi.org/10.20430/ETE.V87I348.1167

Soto, A. y Cerrano, C. (2020). Disyuntivas económicas y políticas de la guerra fría en América Latina. *Humanidades: Revista de la Universidad de Montevideo*, *7*, 9-20. http://dx.doi.org/10.25185/7.1

Suanes, M. y Roca-Sagalés, O. (2015). Inversión extranjera directa, crecimiento económico y desigualdad en América Latina. *El Trimestre Económico*, *82*(327), 675-706. http://www.scielo.org.mx/scielo.php?script=sci_arttext&pid=s2448-718x2015000300675

Trindade, E. P., Farias Hinnig, M. P., Moreira da Costa, E., Sabatini Marques, J., Cid Bastos, R. y Yigitcanlar, T. (2017). Sustainable development of smart cities: A systematic review of the literature. *Journal of Open Innovation: Technology, Market, and Complexity*, *3*(11), 1-14. https://doi.org/10.1186/s40852-017-0063-2

Valenzuela Espinoza, I. (2019). Capacidades humanas, democracia y Estado de bienestar habilitante: Revitalización de la socialdemocracia en América Latina. *Polis*, *52*, 1-20. https://journals.openedition.org/polis/17048

Vernengo, M. (2020). Una nota sobre los bancos centrales en el centro y en la periferia: Estancamiento secular y restricción externa. *Problemas del Desarrollo*, *51*(202), 45-62. https://doi.org/10.22201/IIEC.20078951E.2020.202.69635

Anexo

## Anexo 1. Resumen bibliométrico de artículos revisados (n = 44)

| | TÍTULO | AUTORES | CITAS | CITAS POR AÑO | AÑO | REVISTA | PAIS REVISTA |
|---|---|---|---|---|---|---|---|
| 1 | Desarrollo local y microfinanzas como estrategias de atención a las necesidades sociales: un acercamiento teórico conceptual | Louis Valentin Mballa | 61 | 15,25 | 2017 | *Revista Mexicana de Ciencia Política* | *México* |
| 2 | La nueva teoría desarrollista: una síntesis | Luiz Carlos Bresser-Pereira | 34 | 8,50 | 2017 | *Economía UNAM* | *México* |
| 3 | Inversión extranjera directa, crecimiento económico y desigualdad en América Latina | Macarena Suanes y Oril Roca-Sagalés | 48 | 8,00 | 2015 | *El Trimestre Económico* | *México* |
| 4 | Gasto público y crecimiento económico. Un estudio empírico para América Latina | Diego Pinilla Rodríguez, Juan Jiménez Aguilera y Roberto Montero Granados | 53 | 6,63 | 2013 | *Cuadernos de Economía* | *Colombia* |
| 5 | La infraestructura en el desarrollo de América Latina | Jorge Kogan y Diego Bondorevsky | 28 | 5,60 | 2016 | *Economía y Desarrollo* | *Cuba* |
| 6 | Ciclos distributivos y crecimiento económico en América Latina, 1950-2014 | Germán Alarco Tosoni | 18 | 4,50 | 2017 | *Cuadernos de Economía* | *Colombia* |
| 7 | Inversión extranjera directa y desarrollo en América Latina | Josefina Morales | 45 | 4,09 | 2010 | *Problemas del Desarrollo* | *México* |
| 8 | Volver al desarrollo | Jaime Ornelas Delgado | 36 | 4,00 | 2012 | *Problemas del Desarrollo* | *México* |
| 9 | Unions and economic performance in developing countries: case studies from Latin America | Fernando Rios-Avila | 15 | 3,75 | 2017 | *Ecos de Economía* | *Colombia* |
| 10 | Materias primas críticas y complejidad económica en América Latina | Juan Sebastián Lara-Rodríguez; André Tosi Furtado y Aleix Altimiras-Martin | 11 | 3,67 | 2018 | *Apuntes del Cenes* | *Colombia* |
| 11 | Inversión en infraestructura vial y su impacto en el crecimiento económico: Aproximación de análisis al caso infraestructura en Colombia (1993-2014) | Miguel David Rojas-López y Andrés Felipe Ramírez-Murie | 11 | 3,67 | 2018 | *Revista de Ingeniería Universidad de Medellín* | *Colombia* |
| 12 | Raíces latinoamericanas del otro desarrollo: estilos de desarrollo y desarrollo a escala humana | Rafael Domínguez y Sara Caria | 10 | 3,33 | 2018 | *América Latina en la Historia Económica* | *México* |
| 13 | Mercados accionarios y su relación con la economía real en América Latina | Samuel Brugger y Edgar Ortiz | 30 | 3,33 | 2012 | *Problemas del Desarrollo* | *México* |
| 14 | El desarrollo ayer y hoy: idea y utopía | Rolando Cordera Campos | 21 | 3,00 | 2014 | *Economía UNAM* | *México* |
| 15 | El desarrollo en las ideas de la Cepal | Hugo Beteta y Juan Carlos Moreno-Brid | 25 | 2,78 | 2012 | *Economía UNAM* | *México* |
| 16 | Insurance use and economic growth in Latin America. Some panel data evidence | Rodrigo Taborda y Ángela Concha | 17 | 2,43 | 2014 | *Lecturas de Economía* | *Colombia* |
| 17 | ¿Qué sucedió con el casillero vacío del desarrollo de América Latina 30 años después? | Paola Jaimes y Guillermo Matamoro | 9 | 2,25 | 2017 | *Problemas del Desarrollo* | *México* |
| 18 | Raúl Prebisch y el pensamiento estructuralista latinoamericano | Julio López | 2 | 2,00 | 2020 | *Problemas del Desarrollo* | *México* |
| 19 | Variedades de capitalismo, implicaciones para el desarrollo de América Latina | Mario Humberto Hernández López | 8 | 2,00 | 2017 | *Economía teoría y práctica* | *México* |
| 20 | Implicaciones de la crisis financiera y económica global en América Latina | Susana Nudelsman | 16 | 2,00 | 2013 | *Problemas del Desarrollo* | *México* |
| 21 | La productividad y el desarrollo en América Latina dos interpretaciones | Jaime Ros | 17 | 1,70 | 2011 | *Economía UNAM* | *México* |
| 22 | De los Aspectos Sociales del Desarrollo Económico a la Teoría de la Dependencia: Sobre la gestación de un pensamiento social propio en Latinoamérica | Juan Jesús Morales | 15 | 1,67 | 2012 | *Cinta de Moebio* | *Chile* |
| 23 | Desarrollo económico de América Latina y las integraciones regionales del siglo XXI | Rodolfo Iván González Molina | 11 | 1,22 | 2012 | *Ecos de Economía* | *Colombia* |
| 24 | Una nota sobre los bancos centrales en el centro y en la periferia: estancamiento secular y restricción externa | Matías Vernengo | 1 | 1,00 | 2020 | *Problemas del Desarrollo* | *México* |
| 25 | Evaluación del impacto económico de programas sociales contra la pobreza: una revisión de estudios aleatorizados en la obra de Esther Duflo | Jaiberth Antonio Cardona Arias | 1 | 1,00 | 2020 | *Revista Facultad Nacional de Salud Pública* | *Colombia* |
| 26 | Auditorías de desempeño en América Latina: ¿Mejoran la confianza en los gobiernos? | Ana Yetano y Blanca Isela Castillejos | 2 | 1,00 | 2019 | *Gestión Política Pública* | *México* |
| 27 | Avances fundamentales de la economía feminista en América Latina | Natalia Escobar Váquiro | 4 | 1,00 | 2017 | *Cuadernos de Economía Crítica* | *Argentina* |
| 28 | América Latina bajo el fuego de las grandes transformaciones económicas y políticas | Jaime Osorio | 8 | 0,89 | 2012 | *Política y Cultura* | *México* |

| | | | | | | |
|---|---|---|---|---|---|---|
| 29 | Desigualdades, territorios y vulnerabilidades en el desarrollo de América Latina (1990-2015) | José Ramón Carmona Motolinia y Esther Iglesias Lesaga | 3 | 0,60 | 2016 | *Sociedad y Economía* | *Colombia* |
| 30 | Eficiencia de la innovación en América Latina. Una aproximación a través del Análisis Envolvente de Datos | Antonio Favila Tello | 1 | 0,50 | 2019 | *Análisis Económico* | *México* |
| 31 | Modelos de crecimiento, estacionariedad y rompimientos: comparación entre las tendencias de crecimiento de las economías de la OCDE y las de los países menos desarrollados | Zeus Salvador Hernández-Veleros | 2 | 0,40 | 2016 | *El Trimestre Económico* | *México* |
| 32 | Recetas contra el conjuro: los estilos de desarrollo en el Uruguay contemporáneo y los debates sobre el desarrollo latinoamericano (1973-2014) | Sebastian Levalle | 1 | 0,33 | 2018 | *Dialogos (revista electrónica de historia)* | *Costa Rica* |
| 33 | Auge y caída de la planificación regional en Argentina: entre la búsqueda de una autonomía enraizada y el escenario económico del capitalismo periférico (1965-2015) | Ariel García | 1 | 0,33 | 2018 | *Cuadernos de Geografía: Revista Colombiana de Geografía* | *Colombia* |
| 34 | del vivir bien y del desarrollo sustentable. Extractivismos y construcción de alternativas al desarrollo en Bolivia y chile. | Miriam Lehnert y Noelia Carrasco | 0 | 0,00 | 2020 | *Dialogo Andino* | *Chile* |
| 35 | ¿Por qué los países latinoamericanos sufren un estancamiento económico de largo plazo? Un estudio a partir de los casos de Argentina, Brasil y México | Pierre Salama | 0 | 0,00 | 2020 | *El Trimestre Económico* | *México* |
| 36 | Clave hermenéutica del desarrollismo | Horacio Bossio García | 0 | 0,00 | 2020 | *Revista de Historia (Concepción)* | *Chile* |
| 37 | Impacto de los combustibles fósiles, el consumo de energía renovable y el crecimiento industrial en las emisiones de carbono en las economías de América Latina y el Caribe | Muhammad Khalid Anser, Imran Hanif, Majed Alharthi y Imran Sharif Chaudhr | 0 | 0,00 | 2020 | *Atmosfera* | *México* |
| 38 | Disyuntivas económicas y políticas de la Guerra Fría en América Latina | Ángel Soto y Carolina Cerrano | 0 | 0,00 | 2020 | *Humanidades* | *Uruguay* |
| 39 | Economía Social de Mercado versus capitalismo rentista. Reflexiones para América Latina | Marcelo F. Resico | 0 | 0,00 | 2019 | *Civilizar* | *Colombia* |
| 40 | Análisis de cointegración entre el crecimiento económico y deterioro medio-ambiental. Un análisis empírico del desarrollo sostenible de América Latina y el Caribe | Víctor Quinde Rosales, Gabriela Vaca Pinela, Francisco Quinde Rosales y Lourdes Lazo Vaca | 0 | 0,00 | 2019 | *Revista de Economía del Caribe* | *Colombia* |
| 41 | Capacidades humanas, democracia y Estado de | Iván Valenzuela Espinoza | 0 | 0,00 | 2019 | *Polis* | *Chile* |
| | bienestar habilitante: revitalización de la socialdemocracia en América Latina | | | | | | |
| 42 | Producto interno Bruto en América Latina y el Caribe: Relaciones entre crecimiento económico y sustentabilidad ambiental | Víctor Quinde-Rosales, Rina Bucaram-Leverone y Martha Bucaram-Leverone | 0 | 0,00 | 2019 | *Revista Venezolana de Gerencia* | *Venezuela* |
| 43 | ¿Es procíclica la convergencia del crecimiento económico de América Latina? | María Isabel Osorio Caballero | 0 | 0,00 | 2019 | *Investigación Económica* | *México* |
| 44 | Economic growth and tourism performance in Latin America and the Caribbean: a comparative analysis by clustering techniques and causality tests | Verónica Segarra, Martín Olivera y Juan Gabriel Brida | 0 | 0,00 | 2021 | *Revista Brasileira de Pesquisa em Turismo* | *Brasil* |

## Anexo 2. Resumen de objetivos de artículos seleccionados (n = 13)

| | **Artículo** | **Objetivo** |
|---|---|---|
| 1 | Mballa, Louis Valentin. (2017). Desarrollo local y microfinanzas como estrategias de atención a las necesidades sociales: un acercamiento teórico conceptual. Revista mexicana de ciencias políticas y sociales, 62(229), 101-127 | Explorar las diferentes aproximaciones teóricas a las nociones de desarrollo local y microfinanzas para establecerlas como herramientas de atención a las necesidades socioeconómicas de las personas de escasos recursos. Se parte de la hipótesis de que el desarrollo local y las microfinanzas son, en efecto, instrumentos fundamentales para dar respuesta a las necesidades sociales como causas de la pobreza. |
| 2 | Bresser-Pereira, Luiz Carlos. (2017). La nueva teoría desarrollista: una síntesis. Economía UNAM, 14(40), 48-66. | Un análisis sintético de la historia/evolución de la economía del desarrollo y su estado actual. Presenta un marco de análisis crítico y descriptivo del desarrollo como disciplina e introduce de manera sintética y clara la propuesta del nuevo desarrollismo. |
| 3 | Suanes, Macarena y Roca-Sagalés, Oriol. (2015). Inversión extranjera directa, crecimiento económico y desigualdad en América Latina. El trimestre económico, 82(327), 675-706. | Analiza la relación entre la inversión extranjera directa (IED), el crecimiento económico y la desigualdad de ingresos en América Latina a través de un análisis econométrico, para probar la hipótesis de si la IED tiene efectos no lineales sobre la desigualdad. |
| 4 | Pinilla Rodríguez, Diego, Jiménez Aguilera, Juan y Montero Granados, Roberto (2013). Gasto público y crecimiento económico. Un estudio empírico para América Latina. Cuadernos de Economía, 32(59), 181-210. | Analiza la relación entre gasto público y nivel de producción para un conjunto de países de América Latina entre la década de 1990 y la primera década del siglo XXI, a través de un análisis econométrico, para probar la hipótesis de que la relación entre ambas variables no es lineal. |
| 5 | Kogan, Jorge y Bondorevsky, Diego (2016). La infraestructura en el desarrollo de América Latina. *Economía y desarrollo*, *156*(1), 168-186. | Reflexiona sobre el rol de la infraestructura en el desarrollo, en particular de los mecanismos de inversión entre el sector público y el privado, los requerimientos financieros, el papel de las instituciones y el impacto social y ambiental de los proyectos de infraestructura. |
| 6 | Alarco Tosoni, Germán (2017). Ciclos distributivos y crecimiento económico en América Latina, 1950-2014. *Cuadernos de Economía*, *36*(spe72), 1-42. | Analiza la relación entre participación de los salarios y el crecimiento económico, a través de un ejercicio econométrico en que analiza 18 economías latinoamericanas, y el conjunto regional, para probar la hipótesis de que la participación salarial es positiva para el crecimiento. |
| 7 | Morales, Josefina (2010). Inversión extranjera directa y desarrollo en América Latina. *Problemas del desarrollo*, *41*(163), 141-156. | Reflexiona sobre el rol asignado a la inversión extranjera directa (IED) como factor indispensable para el desarrollo. |
| 8 | Ornelas Delgado, Jaime (2012). Volver al desarrollo. *Problemas del desarrollo*, *43*(168), 7-35. | Reflexiona acerca de la evolución y el estado del desarrollo en América Latina, para contribuir a la construcción de una agenda que permita la deconstrucción del desarrollo. |
| 9 | Rios-Avila, Fernando (2017). Unions and Economic Performance in developing countries: case studies from Latin America. *Ecos de Economía*, *21*(44), 4-36. | Analiza el impacto económico de los sindicatos sobre la productividad en el sector manufacturero, realizando un ejercicio econométrico con datos de seis países latinoamericanos. |
| 10 | Lara-Rodríguez, Juan Sebastian, Furtado, André Tosi y Altimiras-Martin, Aleix (2018). Materias primas críticas y complejidad económica en América Latina. *Apuntes del CENES*, *37*(65), 15-51. | Examina las políticas minerales de materias primas críticas en las principales economías de América Latina y el papel de sus respectivos sistemas nacionales de innovación (SNI), en búsqueda de mayor complejidad económica, mediante un método de evaluación comparativo aplicado a la política mineral de seis países de la región. |
| 11 | Rojas-López, Miguel David y Ramírez-Muriel, Andrés Felipe (2018). Inversión en infraestructura vial y su impacto en el crecimiento económico: Aproximación de análisis al caso infraestructura en Colombia (1993-2014). Revista Ingenierías Universidad de Medellín, 17(32), 109-128 | Analiza la evolución y la relación entre inversión en infraestructura vial y el crecimiento económico de Colombia, realizando un ejercicio econométrico en que se compara a Colombia con algunos países de América Latina, para probar la hipótesis de que la relación es positiva. |
| 12 | Brugger, Samuel y Ortiz, Edgar (2012). Mercados accionarios y su relación con la economía real en América Latina. *Problemas del desarrollo*, *43*(168), 63-93. | Examina la relación entre el desempeño de las bolsas de valores de cuatro países latinoamericanos con la evolución de su economía real, aplicando siete modelos econométricos de series de tiempo, para probar la hipótesis de que se afectan mutuamente. |
| 13 | Domínguez, Rafael y Caria, Sara. (2018). Raíces latinoamericanas del otro desarrollo: estilos de desarrollo y desarrollo a escala humana. *América Latina en la historia económica*, *25*(2), 175-209 | Realizan una reconstrucción histórica de las ideas del desarrollo alternativo en América Latina para probar la hipótesis de que estas están conectadas (latentemente) con las variantes actuales de los estilos del desarrollo. |